# WARM *SPITZER* OBSERVATIONS OF THREE HOT EXOPLANETS: XO-4b, HAT-P-6b AND HAT-P-8b

Kamen O. Todorov[1], Drake Deming[2,3], Heather A. Knutson[4,5], Adam Burrows[6], Pedro V. Sada[7], Nicolas B. Cowan[8], Eric Agol[9], Jean-Michel Desert[10], Jonathan J. Fortney[11], David Charbonneau[10], Gregory Laughlin[11], Jonathan Langton[12], Adam P. Showman[13], Nikole K. Lewis[13]

*Draft version November 24, 2011*

## ABSTRACT

We have analyzed Warm *Spitzer/IRAC* observations of the secondary eclipses of three planets, XO-4b, HAT-P-6b and HAT-P-8b. We measure secondary eclipse amplitudes at $3.6\,\mu$m and $4.5\,\mu$m for each target. XO-4b exhibits a stronger eclipse depth at $4.5\,\mu$m than at $3.6\,\mu$m, which is consistent with the presence of a temperature inversion. HAT-P-8b shows a stronger eclipse amplitude at $3.6\,\mu$m, and is best-described by models without a temperature inversion. The eclipse depths of HAT-P-6b can be fitted with models with a small or no temperature inversion. We consider our results in the context of a postulated relationship between stellar activity and temperature inversions and a relationship between irradiation level and planet dayside temperature, as discussed by Knutson et al. (2010) and Cowan & Agol (2011), respectively. Our results are consistent with these hypotheses, but do not significantly strengthen them. To measure accurate secondary eclipse central phases, we require accurate ephemerides. We obtain primary transit observations and supplement them with publicly available observations to update the orbital ephemerides of the three planets. Based on the secondary eclipse timing, we set upper boundaries for $e\cos(\omega)$ for HAT-P-6b, HAT-P-8b and XO-4b and find that the values are consistent with circular orbits.

*Subject headings:* stars: planetary systems — eclipses – techniques: photometric

## 1. INTRODUCTION

The *Spitzer Space Telescope* has enabled direct measurements of light emitted by exoplanets known as "hot Jupiters" through time series photometry during secondary eclipse. The hot Jupiters have masses and radii comparable to the gas giants in the Solar System, but their orbital semi-major axes are very small, with periods between 1 and 5 days, and equilibrium temperatures over 1000 K. The wavelength dependent drop of total light from the star-planet system as the planet moves behind the star during a secondary eclipse was first measured independently by Charbonneau et al. (2005) and Deming et al. (2005). Measuring the eclipse depth at multiple wavelengths allows the construction of very low resolution, infra-red emergent spectra of the day side of the planet (Charbonneau et al. 2008; Grillmair et al. 2008). Comparison of these measurements to models suggests that there are two subclasses of hot Jupiters, based on the presence or absence of strong temperature inversions in the upper layers of their atmospheres (e.g., Knutson et al. 2008; Machalek et al. 2009; Todorov et al. 2010; Beerer et al. 2011; Deming et al. 2011). The reason for such inversions is poorly understood, but it is generally believed that planets with a temperature inversion have an additional opacity source in the upper atmosphere. This source of opacity was initially suggested to be gas phase TiO (Hubeny et al. 2003; Burrows et al. 2007, 2008; Fortney et al. 2006b, 2008). However, TiO may be lost to cold traps in the deeper atmospheres and night sides of planets such as HD 209458b where temperatures are predicted to cross below the condensation point for this molecule (Spiegel et al. 2009). Moreover, while TrES-3 is hot enough to have gas phase TiO, it seems to lack a temperature inversion (Fressin et al. 2010). In contrast, XO-1b is too cool for gas-phase TiO, and yet appears to have an inversion (Machalek et al. 2008). In light of these discrepancies, alternate opacity sources may be needed to explain the full range of observations. More recently Zahnle et al. (2009) have suggested that sulfur compounds may be responsible for the high altitude opacity in some hot Jupiter atmospheres. The presence of temperature inversions may also be related to the magnetic activity and corresponding UV flux of the host star (Knutson et al. 2010). According to this hypothesis, active stars are orbited by hot Jupiters that have no temperature inversion in their atmospheres, while the planets around quiet stars have inverted atmospheres.

Hot Jupiters are thought to become tidally evolved in less than $\sim 1\,$Gyr, and, assuming zero eccentricity, their rotation periods should be equal to their orbital periods (Correia & Laskar 2010). Heat redistribution from the day side to the night side is an important process that has an impact on the emergent day-side spectrum of the planet. The heat redistribution efficiency influences the dayside temperature, which in turn is one of the factors that control the secondary eclipse depth. Comparing observed secondary eclipse depths with atmospheric models can constrain the redistribution efficiency and Bond albedo of the planet's atmosphere, since we know the total amount of energy the planet receives from its host star.

[1] Department of Astronomy and Astrophysics, The Pennsylvania State University, University Park, PA 16802
[2] Planetary Systems Laboratory, NASA's Goddard Space Flight Center, Greenbelt, MD 20771
[3] Present address: Department of Astronomy, University of Maryland at College park, College Park, MD, 20742
[4] Division of Geological and Planetary Sciences, California Institute of Technology, Pasadena, CA 91125
[5] University of California Berkeley, Department of Astronomy, Berkeley, CA 94720
[6] Department of Astrophysical Sciences, Princeton University, Princeton, NJ 05844
[7] University of Monterrey, Monterrey, Mexico
[8] CIERA Fellow, Department of Physics & Astronomy, Northwestern University, Evanston, IL 60208
[9] Department of Astronomy, University of Washington, Box 351580, Seattle, WA 98195
[10] Harvard-Smithsonian Center for Astrophysics, Cambridge, MA 02138
[11] Department of Astronomy and Astrophysics, University of California at Santa Cruz, Santa Cruz, CA 95064
[12] Department of Physics, Principia College, Elsah, IL 62028
[13] Lunar and Planetary Laboratory, University of Arizona, Tucson, AZ 85721

After the cryogen on *Spitzer* ran out in May 2009, the observatory has continued to work at a higher temperature of roughly 29 K, cooled by passive radiation. During the warm phase of the mission observations at only two pass-bands are possible – only the 3.6 and 4.5 $\mu$m channels of the Infra-Red Array Camera (IRAC, Fazio et al. 2004) are still operational. However, measurements at these two wavelengths are often sufficient to place constraints on the presence or absence of strong temperature inversions in these atmospheres.

In this paper we present *Warm Spitzer* time series photometry, in both 3.6 and 4.5 $\mu$m, for three hot Jupiters – XO-4b (McCullough et al. 2008), HAT-P-6b (Noyes et al. 2008) and HAT-P-8b (Latham et al. 2009). The properties of these planets and their host stars that we adopt are listed in Table 1. We derive secondary eclipse depths for each planet and place constraints on the properties of their dayside atmospheres. In addition, we combine available data on the primary transits of these planets with our own ground based primary transit observations and update the available ephemerides. We place upper bounds on the orbital eccentricities of the planets, based on the secondary eclipse timing. As expected for objects on tight orbits, the measurements are consistent with circular orbits.

In Section 2 we discuss the *Spitzer* images and their properties, our photometry technique and our ground based observations. Our time-series analysis procedure and uncertainty estimation are presented in Section 3. We discuss our results in the context of previous studies in Section 4.

## 2. OBSERVATIONS AND PHOTOMETRY

### 2.1. *Secondary Eclipse Observations with Spitzer*

All observations were performed with the IRAC instrument in subarray mode. The images are $32 \times 32$ pixels in size, corresponding to $39'' \times 39''$, centered on the planet's host star. Subarray mode observations result in FITS data cubes each containing 64 images taken consecutively. Observations were made in both the 3.6 $\mu$m and the 4.5 $\mu$m channels during different secondary eclipses, with effective exposure times per image of 1.92 s, for both observing wavelengths. At each wavelength, the HAT-P-6 and HAT-P-8 observations lasted for 461 min, resulting in 13,760 images (215 data cubes), while the XO-4 observations lasted for 472 min per wavelength (14,080 images, or 220 data cubes). Information about the time span of the observations is presented in Table 2.

#### 2.1.1. *Photometry Extraction*

We extract photometry from the Basic Calibrated Data (BCD) files produced by version S18.18.0 of the Spitzer pipeline. We calculate the orbital phase of the planet at the middle of a given observation based on the MJD_OBS keyword in the FITS headers. This time is given in modified Julian date, based on the Coordinated Universal Time (UTC) standard. Following the discussion in Eastman et al. (2010), UTC is based on the International Atomic Time standard, which uses the hyperfine transitions in caesium 133 atoms. But UTC is discontinuous with leap seconds introduced roughly every year in order to keep it within 0.9 seconds from UT1, which is based on the mean solar day. The Terrestrial Time (TT) standard is continuous and offset from UTC by $32.184 \, \text{sec} + \text{N}$, where N is the current number of leap seconds. For our Spitzer observations, $\text{N} = 34$ seconds. Barycentric Dynamic Time (TDB) is similar to TT but includes a relativistic correction. The size of the correction is typically few milliseconds, and for our purposes it is negligible. We convert the $\text{MJD}_\text{UTC}$ (modified Julian date in the UTC standard) times from the *Spitzer* FITS file headers to barycentric Julian date based on the TDB standard ($\text{BJD}_\text{TDB}$) using Jason Eastman's IDL routine *get_spitzer_bjd*.

In order to facilitate the estimation of photometric errors, we convert the pixel intensities to electron counts using the calibration information from the image headers, immediately after reading the images. To remove the background radiation from the images, we create a histogram of the pixel values for each frame, which we fit with a Gaussian. We exclude from the histogram the values coming from the $5 \times 5$ pixel square centered on the star to avoid biasing the background measurement towards higher values. The location of the Gaussian peak gives us an estimate of the background radiation level, and we subtract it from each pixel in the image. Pixel values that are more than $4\sigma$ away from a boxcar median of width 5 of the same pixel through time within each data cube are flagged as energetic particle hits. These pixels are corrected by replacing their value with the boxcar median. The corrected pixels are 0.45-0.50% of all pixels at 3.6 $\mu$m and 0.11-0.13% at 4.5 $\mu$m.

Our photometric routines locate the centroid of the stellar point spread function (PSF) by fitting a symmetric two-dimensional Gaussian function to the PSF core (Agol et al. 2010). We measure circular aperture photometry, with varying radii in increments of 0.5 pixels between 1.5 and 4.0 pixels. The measured eclipse depths for all planets and wavelengths varied by less than $1\sigma$ when varying the photometry aperture radius. The best aperture radius is selected based on the scatter that it produces in the photometry after the best-fit solution for the detector effects has been removed. We find that the scatter is weakly dependent on the aperture radius, with least scatter found at 3.0 (HAT-P-8 at 3.6 $\mu$m) and 2.5 (all other data) pixel radii. The raw photometric light curves for the three planets are presented in Figures 1 (3.6 $\mu$m data) and 2 (4.5 $\mu$m data).

### 2.2. *Ground-Based Transit Observations*

The phase of the secondary eclipse is sensitive to the eccentricity of the planetary orbit. Although the orbits of close-in giant planets are expected to be circularized by tidal forces within $\sim 1 \, \text{Gyr}$ (Correia & Laskar 2010), residual eccentricity can persist in a few cases and those small eccentricities can provide important insights into the dynamical evolution of exoplanetary systems. In order to exploit the secondary eclipse timings derived from our Spitzer data, we must have accurate and precise ephemerides for the planetary transits. Unfortunately, we are aware of only a few follow-up transit observations for HAT-P-6, HAT-P-8, and XO-4 since their discovery (Narita et al. 2010; Szabo et al. 2010). A resource for observations of exoplanet transits is the Czech Astronomical Society's Transiting Exoplanets and Candidates (TRESCA) project and database[14], where observers can upload transit light curves and report the best fit parameters they extract. The primary transit observations reported in the TRESCA database are often of poor quality, making them unsuitable for ephemeris measurements. However, we include in our analysis two transit observations in the R band available in TRESCA, by Ramon Naves (HAT-P-6b, 12" telescope, Montcabrer, Spain) and by Felix Hormuth (HAT-P-8b, 1.2m telescope, Calar Alto Observatory, Spain). These observations have clean photometry and relatively small measurement uncertainties. We adopt the published transit timing values for all of the above measurements.

We observed additional transits of these systems using the Universidad de Monterrey Observatory (UDEM) telescope on 5 occasions between 1 October 2008, and 1 August 2010. UDEM is a small private college observatory having Minor Planet Center Code 720, and located at 689 meters altitude in the suburbs of Monterrey, México (Sada et al. 2010). The UDEM data were acquired using an $I_c$-band filter on the 0.36-meter reflector, with a 1280x1024-pixel CCD camera at $1''\!.0$ pixel$^{-1}$ scale. The CCD exposures were binned 2x2 to facilitate rapid readout. We observed an additional transit of HAT-P-6 on 25 November 2009 (UT) using the 2.1-meter reflector on Kitt Peak, with the FLAMINGOS 2048x2048-pixel infrared imager (Elston et al. 1998), and a J-band (1.25 $\mu$m) filter, at 0.6 arc-sec per pixel scale. Observations at both observatories used a defocus to improve the photometric precision, and both used off-axis guiding to maintain pointing stability. The exposures for each image are 40 s for XO-4, and 30 s for the HAT planets, the photometric aperture used is between $3''\!.9$ and $5''\!.9$, and we used between 3 and 8 comparison stars for each observation. The detrended and normalized light curves of the transits we observed are shown in Figure 3. Our detrending procedure is described in Section 3.1. Flat-field observations were acquired at both observatories using either twilight sky (UDEM) or a series of night-sky exposures (FLAMINGOS), incorporating pointing offsets to allow removal of stars via a median filter.

## 3. DATA ANALYSIS

### 3.1. *Improved Ephemeris Estimates*

In order to estimate the central phases and their uncertainties in the secondary eclipse fits accurately, we first need to have precise and accurate ephemerides. Using the ground-based observations described in Section 2.2, we perform aperture photometry on the three transit stars and several comparison stars. Comparison stars are added and deleted as necessary in order to achieve the best photometric precision. Similarly, the aperture radius for the photometry is varied to achieve best precision. The central transit times are derived by fitting a transit curve based on the parameters of the systems as announced in their discovery papers. We shift the transit curves in time, and scale them in depth if necessary, to fit the observed photometry. In the FLAMINGOS case the fit also requires a polynomial baseline. The baseline curvature is likely due to differences in spectral type between the target and comparison stars, combined with the wavelength dependence of telluric water absorption (Sada et al. 2010). We estimate the error of the fitted transit time as the quadratic sum of two independent contributions. The first contribution is caused by the scatter of the data around the transit curve, after having removed the polynomial baseline. We calculate this uncertainty in time by dividing the difference in intensity between the data and the best fit transit curve (i.e., the random scatter) of the individual points by the slope of the best fit transit curve at each point, and then integrating the result over time. The second contribution is due to imprecision in the baseline, and we estimate this error by a bootstrap Monte Carlo process, creating multiple realizations of the baseline and calculating how their different slopes across the transit curve would affect the fitted central time. The baseline contribution to uncertainty is small – on the order of a second or less compared to roughly a minute from scatter for the UDEM observations and 30 s for the KPNO observation. Central transit times, their total uncertainties and resulting ephemeris are included in Table 3. The offsets in minutes of the observed transits from the best fit ephemerides are shown in Figure 4. Due to the new transit timings, the contribution of the ephemerides uncertainties to the secondary eclipse time uncertainties for our data sets is reduced from ∼67 to ∼65 seconds for XO-4b, from ∼160 to ∼49 seconds for HAT-P-6b, and from ∼91s to ∼44s for HAT-P-8b.

[14] http://var2.astro.cz/EN/tresca/

### 3.2. *Secondary Eclipse Fits*

In order to measure the depth of the secondary eclipse we need to correct the *Spitzer* photometry for instrumental effects. We start the analysis by normalizing the data to unity for the time when the light from the star only is visible (secondary eclipse). We proceed by removing data points that have been measured to have high backgrounds or are outliers in the XO-4 observations – 13 photometric points in the 3.6 $\mu$m data and 5 in the 4.5 $\mu$m data. This operation is not necessary for the other data sets. In addition, the 57$^{\text{th}}$ frame in each data cube at 4.5 $\mu$m exhibits a relatively high background value. All of these frames are removed from the analysis (215-220 points per data set). Photometry data points coming from images with high background are often not outliers, particularly the 57$^{\text{th}}$ frames. Still, we choose to omit them from the analysis, since the high background indicates the possible presence of systematic effects in these particular images that we cannot account for. The images in the 29$^{\text{th}}$ data cube of the HAT-P-8 observation at 4.5 $\mu$m exhibit values of their brightest stellar pixel four orders of magnitude higher than the median of this pixel in the rest of the data cubes, and were excluded (64 images).

Previous experience shows that two different transient effects are seen in *Spitzer/IRAC* data, for both the cryogenic and Warm mission. First, there is often an initial position instability wherein it takes tens of minutes for the star's position to stabilize on the detector (e.g., Anderson et al. 2011). That can cause a perturbation in intensity because it involves portions of the detector that have different intrapixel sensitivity variations than the ones used in most of the observation. A second effect has been seen by some investigators (e.g., Campo et al. 2011; Deming et al. 2011) wherein the initial portion of the $3.6\,\mu$m data shows a "ramp" effect that increases or decreases the intensity with a small exponential-like behavior, and is uncorrelated with pixel position. This initial ramp quickly asymptotes to a constant level, within the noise, so the simplest method to account for it is to clip the first portion of the data. Therefore, we also omit the first 466 points (phase less than 0.455, or 15.7 min of data) in the XO-4 data at $3.6\,\mu$m since they show a steep upward ramp, with 6% change in measured flux. In the HAT-P-8 light curve at $3.6\,\mu$m, the photometric values at the beginning of the observation are $\sim 1-2\%$ higher than average, which is an effect not correlated with pixel position and is therefore difficult to model. We reject the first 1056 points (first 35.5 min of data, with phase less than 0.445) of this data set. The first 1920 photometric points for HAT-P-6 at $3.6\,\mu$m (first 65 min, with phase less than 0.465) are rejected because of an upward ramp-like behaviour and because we note a sharp peak in brightness at phase of $\sim 0.4635$, which lasts for roughly 10-15 minutes and has an amplitude of $\sim 0.3\%$ above the local average in the binned data after the correction for instrumental effects. A similar effect was seen by Stevenson et al. (2010) in their *Spitzer/IRAC* $3.6\,\mu$m photometric time series during a secondary eclipse of GJ 436b. If not for this transient effect, clearly unrelated to the eclipse, we would have rejected only the first approximately 30min of data on HAT-P-6 at $3.6\,\mu$m. We also reject the first 876 data points for HAT-P-6b at $4.5\,\mu$m, corresponding to 30 min of data with phase less than 0.45822. These points are poorly accounted for by our secondary eclipse and instrumental effects models and cause significant increases in RMS residual scatter of the data, if included in the analysis. We determine the number of points to be rejected by visual examination of the residual light curves and noting that at the start of these data sets there are systematic deviations from the best fit model. We then estimate the time when the systematic effects begin to be well accounted for and remove all points before this moment. Removing the initial data points from the HAT-P-8 and XO-4 at $4.5\,\mu$m is not necessary since the instrumental effects are modeled accurately. Therefore, we choose not to clip these data.

To further clean the data from potential outliers, we subtract the boxcar median of width 9 of the photometric data from the signal, removing any data points with residuals $4\sigma$ or more away from zero. The number of points removed in this way is between 5 and 20 points, varying between different wavelengths and planets. Removing so few outlying points has no effect on the measurements of the eclipse depths and central phases. The total fraction of photometric points removed from the analysis is between 1.6% and 13.9% for the different data sets.

Similarly to e.g., Knutson et al. (2009), Todorov et al. (2010) and Beerer et al. (2011), we find that there is a dependence between the position of the stellar centroid on its array pixel in both X and Y and the measured brightness of the star. The only two data sets in which the stellar centroid crosses from one pixel to the next are XO-4 and HAT-P-6 at $4.5\,\mu$m in the Y and X directions, respectively. We assume that the two neighboring pixels have similar dependencies of sensitivity on X and Y. When we decorrelate the measured intensity versus the X and Y position of the star, we notice that the eclipse depth depends on the choice between linear and quadratic dependence within $0.5\sigma$ for HAT-P-8 and XO-4 at $4.5\,\mu$m and HAT-P-6 at $3.6\,\mu$m. HAT-P-8 at $3.6\,\mu$m changes upward by $1.3\sigma$. More significantly, the eclipse depth of XO-4 at $3.6\,\mu$m changes upward by $3.7\sigma$, and the eclipse depth of HAT-P-6 at $4.5\,\mu$m changes also upward by $2.7\sigma$. We calculated the reduced $\chi^2$ values and the Bayesian Information Criterion (BIC, Stevenson et al. 2010) for all eclipse fits, varying the greatest exponent of X and Y-pixel positions from 0 to 4. Considering the totality of the BIC values as a function of the exponent, we find a broad minimum near exponent 2. There is sufficient scatter in the BIC values that we could not rigorously defend this choice for all individual eclipses. Nevertheless, the totality of our results indicates that quadratic functions of X and Y are the best choice to model the intrapixel effect. In order to produce consistent results, we adopt the quadratic fit for all data sets. We note that a quadratic decorrelation versus X and Y is consistent with the conclusions of many previous investigations (e.g., Charbonneau et al. 2008; Knutson et al. 2008; Christiansen et al. 2010; Anderson et al. 2011; Cochran et al. 2011; Demory et al. 2011; Desert et al. 2011). In the most extreme case, XO-4 at $3.6\,\mu$m, the greater eclipse depth produced by a fit that is only linear in X and Y will still indicate an inverted atmospheric structure, albeit not so extreme as for the case of our adopted quadratic fit. In the other cases, the change of measured eclipse depth has no effect on the conclusions. Additionally, in the code, we also provide for the possibility of a linear ramp of brightness with time, which has been observed in previous studies (Knutson et al. 2009; Todorov et al. 2010).

We assume an initial eclipse central phase and employ a simultaneous linear regression fit for all instrumental effects described above and for the eclipse. After making an incremental increase of the assumed central phase, we repeat the linear regression measurement. The phase step size is $10^{-5}$ for all data sets, and we cover the phase interval between 0.49 and 0.51. As we step through central phases, we calculate the $\chi^2$ goodness of the fit to the data, and we take the best fit central phase and measured parameters to be the ones that produce the smallest $\chi^2$ value. In this algorithm, the parameters used to model the data are the central phase, the eclipse depth, the linear ramp coefficients, and the coefficients for the pixel sensitivity variation described above. The duration of the eclipse is fixed equal to the duration of the primary transit. We present the best fit parameters that we find in Table 4 and Figures 5 and 6.

Since the measured slopes of brightness with time are close to zero, we experiment by excluding the ramps from the analysis. The measured best eclipse depths without ramps are 0.043% (XO-4b at $3.6\,\mu$m), 0.130% (XO-

4b at 4.5 μm), 0.120% (HAT-P-6b at 3.6 μm), 0.106% (HAT-P-6b at 4.5 μm), 0.132% (HAT-P-8b at 3.6 μm) and 0.127% (HAT-P-8b at 4.5 μm). Lowering the order of the intrapixel variation correction polynomial increases the RMS scatter more that the removal of the linear ramp from the calculations. In fact, the removal of the linear ramp has negligible effect on the scatter and eclipse depths for the data on XO-4 at 4.5 μm (less than 1 σ change in eclipse depth) and HAT-P-6 at 4.5 μm (no change in eclipse depth). A noticeable upward ramp appears in the HAT-P-6 corrected light curve at 3.6 μm, if the linear ramp is removed from the fitting procedure, with an RMS scatter increase similar in magnitude to that of the other 3.6 μm data sets and the HAT-P-8 at 4.5 μm data set. Therefore, we consider the linear ramp to be a small but significant systematic effect, which needs to be accounted for in our data, and we elect to keep the linear ramp slope as a free parameter of the fitting routine in the final analysis for all data sets.

### 3.2.1. *Uncertainty Estimates*

We use two methods to estimate the uncertainty in our measurements of central phase and eclipse amplitude. In both approaches we simulate data and fit an eclipse curve to it the same way we would to real data. In the first method, we simulate data by taking the best fit parameters from the real data, and computing a best fit model for the observed photometry. We then subtract this model from the data and after scrambling the residuals randomly we add them back to the best fit model, achieving a simulated data set (bootstrap Monte Carlo). We create 10,000 simulated data sets for each planet in each wavelength. The fitting algorithm described in Section 3.2 is applied to the simulated data set, and the resulting best fit parameters are recorded and used to estimate the dispersion in the best fit parameters to the observed data.

The second method we employ to estimate the uncertainties in central phase and eclipse depth is often referred to as "prayer bead", described by e.g., Gillon et al. (2007). It is similar to the method above, except instead of scrambling the residuals randomly, after each iteration we take the first residual and make it last, thus moving every other residual one step closer to the first position, like beads in a rosary. We repeat this operation as many times as we have data points, i.e. until each "bead" has completed a full revolution (between 11,800 and 13,900 iterations, for the different data sets). This method has the advantage of preserving information on possible red noise in the data, which would be lost during a Monte Carlo simulation. The resulting parameter values have non-Gaussian distributions, so we report the "1 σ" uncertainty of a parameter as half the range that covers 68% of the simulated data measurements, centered on the best fit value from the original data. The eclipse depth distributions are presented in Figures 7 and 8. The central phase distributions are not shown, since they have similar non-Gaussian morphology to the eclipse depth distributions. We choose to report the uncertainties estimated with the "prayer bead" method, since they include the effects of any red noise in the original data.

The measurement error is not the only source of uncertainty for the central phase determination. Others are the planets' period and "zeroth" primary transit time ($T_0$) uncertainties. Our observations were made 367 and 369 (HAT-P-6b), 250 and 251 (HAT-P-8b) and 166 and 168 (XO-4b) planetary orbits after $T_0$, amplifying the otherwise small uncertainty in the period. To mitigate this problem, we derive more precise ephemerides based on the list of available primary transit measurements discussed in Section 2.2, and listed in Table 3. We add the ephemeris uncertainties quadratically to the central phase measurement uncertainties in our final estimates. The central eclipse $BJD_{TDB}$ timing uncertainties are not influenced by the uncertainties in the ephemerides, and so we only include the measurement uncertainties when reporting the secondary eclipse dates in Table 4. In addition, we have assumed that current uncertainties in the planet's eclipse duration, impact parameter, and the stellar radius have a negligible effect on the best-fit eclipse depths and times.

## 4. DISCUSSION
### 4.1. *Eclipse Amplitudes*

In order to understand what our measurements imply about the planetary atmospheres, we compare models by Burrows et al. (2007, 2008) and Fortney et al. (2005, 2006a, 2008) to the measured eclipse depths, as shown in Figures 9 and 10. Comparing our results to different models is a way to estimate the model dependency of our results. The parameters in the Burrows models that are interesting to us are $\kappa_{abs}$, the absorption coefficient of the unknown stratospheric absorber, and the redistribution parameter, $P_n$, which varies between 0 (no redistribution) and 0.5 (complete redistribution). In Figure 9, the Burrows models that we adopt have $P_n = 0.35$, 0.3 and 0.1 and $\kappa_{abs} = 0.4$, 0.1 and $0\,\mathrm{cm^2 g^{-1}}$, respectively for XO-4b, HAT-P-6b and HAT-P-8b. These values imply a strong temperature inversion for XO-4b, a weak inversion for HAT-P-6b and no inversion for HAT-P-8b. The Fortney models have fewer free parameters than the Burrows models (essentially, only heat redistribution efficiency is a free parameter), and use TiO and VO at equilibrium abundances as the high altitude absorbers, causing the temperature inversion. In contrast, the Burrows models use a generic parametrized absorber at high altitudes. The increased degrees of freedom in the Burrows models can sometimes produce better fits to the measured secondary eclipse depths than the models discussed by Fortney et al. (2005, 2006a, 2008), as in the case of XO-4b shown here. The Fortney models use a different redistribution parametrization. The factor $f$ varies between $f = 0.25$, meaning that flux is evenly redistributed throughout the whole planet, and $f = 0.67$, where no heat flows at all even between dayside regions of different temperature. At $f = 0.5$, the value where heat is evenly redistributed across the dayside, but no heat is transferred to the night side of the planet (Fortney et al. 2008). In Figure 10, we show Fortney models we adopt with $f = 0.5$ (XO-4b inverted, and HAT-P-8b non-inverted) and 0.63 (HAT-P-6b, non-inverted). Stratospheric absorption from TiO/VO is present in the XO-4b model (inverted atmosphere) and removed from the HAT-P-6b and HAT-P-8b models (no temperature inversion).

It is impractical to "fit" the models to the data in the mathematical sense. The free parameters in the Fortney models are $f$ and the presence or absence of TiO in the atmosphere (which is essentially Boolean), while the Burrows models have $\kappa_{\rm abs}$ and $P_n$, but both sets of models lack a mechanism for adjusting those parameters by reference to data points as part of the model calculation itself. The best that can be done with current computational machinery is to generate a series of models with a variety of parameter values, and manually choose the one that provides the best account of the data. In order to choose the models we adopt, we visually examine Burrows models with $\kappa_{\rm abs}$ ranging from 0 (non-inverted) to $0.1\,{\rm cm}^2{\rm g}^{-1}$ (inverted) for HAT-P-6b and HAT-P-8b, and from 0 to $0.4\,{\rm cm}^2{\rm g}^{-1}$ (strongly inverted) for XO-4b. We vary $P_n$ between 0.1 and 0.3 for HAT-P-6b and HAT-P-8b, and between 0.1 to 0.5 for XO-4. The explored Fortney models include or not TiO/VO in the upper atmosphere and have $f$-values ranging from 0.25 to 0.67, although we find that models with $f = 0.5$ generally apply to our results, except in the case of HAT-P-6b, where a value of 0.63 appears to be a better match.

A caveat for both sets of models is that they assume solar composition atmospheres with equilibrium chemistry, except for the Fortney HAT-P-8b non-inverted model, which has metallicity 10 times higher than solar. If the actual compositions differ significantly from these assumptions, the inferred pressure-temperature profiles for these planets may be correspondingly unreliable. Keeping this in mind, we find that while HAT-P-8b has no temperature inversion in the upper layers of its atmosphere, HAT-P-6b has a moderate to no temperature inversion, and XO-4b has a strongly inverted atmosphere. In addition, we calculate the empirical inversion index described in Knutson et al. (2010) for all three planets. We fit the observed planet-star contrasts with blackbody functions for the planets, with freely varying temperatures. The stellar fluxes were taken from Kurucz models appropriate for the given star's temperature (Kurucz 1979). We then subtract the slope of the blackbody curve across the 3.6 and $4.5\,\mu$m IRAC bands from the measured slope across these bands. Our calculations suggest that the index of XO-4b is $0.057\% \pm 0.016\%$, that of HAT-P-6b is $-0.046\% \pm 0.011\%$, and of HAT-P-8b is $-0.060\% \pm 0.015\%$. The uncertainties are just the uncertainties of the eclipse depth measurements added in quadrature and divided by $(4.5-3.6)\,\mu$m for each planet. We have not taken into account the uncertainties in the blackbody planet slopes. These calculations are in agreement with the suggestion by Knutson et al. (2010) that planets with temperature inversions have indices larger than $-0.05\%$, while non-inverted atmospheres have indices with smaller values.

The reasons for the presence or absence of temperature inversions are not completely understood, but it has been suggested that chromospherically active stars tend to have planets with no temperature inversions, while quiet stars tend to have hot Jupiters with inverted temperature profiles (Knutson et al. 2010). We have plotted our results over data from Knutson et al. (2010) in Figure 11. The chromospheric Ca II H & K activity index is calibrated for $B-V > 0.5$, corresponding to stellar effective temperature of less than 6200 K (Noyes et al. 1984). Therefore, the chromospheric activity of XO-4 and HAT-P-6, measured via the Ca II H & K activity index, is not well constrained. For instance, HAT-P-6 has been measured to have $\log({\rm R}'_{\rm HK}) = -5.03 \pm 0.10$ (Hébrard et al. 2011), compared to $-4.799$ seen by Knutson et al. (2010). A smaller activity index would move HAT-P-6 downward in Figure 11, in agreement with the general trend that less active stars have planets with inverted atmospheres, assuming that the Burrows model with a weak inversion is correct. It is also possible that HAT-P-6b has no inversion instead of a weak one, also bringing it to agreement with the trend. HAT-P-8 is close to the calibration limit – it has been measured to have $T_{eff} = 6130$ K Knutson et al. (2010) and 6200 K (Latham et al. 2009). This makes the Ca II H & K activity index for HAT-P-8 somewhat uncertain.

None of the three stars, HAT-P-6, HAT-P-8 or XO-4, is particularly young or old, or rapidly rotating (McCullough et al. 2008; Mamajek & Hillenbrand 2008; Noyes et al. 2008; Latham et al. 2009; Schlaufman 2010; Hébrard et al. 2011). The reported stellar ages are $2.1 \pm 0.6$ (XO-4), $2.3^{+0.5}_{-0.7}$ (HAT-P-6) and $3.4 \pm 1.0$ (HAT-P-8) Gyr, with rotational speeds of $v_{\rm rot}\sin i \approx 8.8, 8.5$ and $11.5\,{\rm km\,s}^{-1}$, respectively, providing little information about the stars' magnetic activity. Therefore, we conclude that the newly measured planets do not contradict the hypothesis of Knutson et al. (2010), but they do not strengthen it either.

Recently, Cowan & Agol (2011) have examined the heat recirculation efficiency of hot Jupiters, its degeneracy with the Bond albedo and its dependence on $T_{\varepsilon=0}$ (the effective temperature of the planet's day side assuming no redistribution and zero Bond albedo). Here, $\varepsilon$ is a recirculation efficiency. For a discussion on the various heat redistribution parametrizations, see e.g., the Appendix to Spiegel & Burrows (2010).

Following the approach described in Cowan & Agol (2011), we calculate brightness temperature of the planet as a function of wavelength, $T_b(\lambda)$, $T_{\varepsilon=0}$, $T_0$ (equilibrium temperature at the substellar point) and $T_d$ (effective dayside temperature) for HAT-P-6b, HAT-P-8b and XO-4b and present the results in Table 5. In order to get the uncertainties in $T_d$, we have averaged the upper and lower uncertainty estimates of the eclipse depth measurement. We find that HAT-P-6b and XO-4b have very similar values for $T_{\varepsilon=0}$, which is expected, given that they have similar host stars and orbital properties. HAT-P-8b orbits the least hot star of the three, but also has the shortest semi-major axis. Therefore, it has the largest value for $T_{\varepsilon=0}$. Still, its effective dayside temperature is similar to that of HAT-P-6b, while XO-4b seems to have a somewhat cooler day side. XO-4b, HAT-P-6b and HAT-P-8b have too low $T_{\varepsilon=0}$ to test the hypothesis by Cowan & Agol (2011) that planets with $T_{\varepsilon=0} \gtrsim 2400$ K have a narrow distribution range of ratios of $T_d / T_0$.

### 4.2. Orbital Phase

A measurement of the secondary eclipse central phase can constrain the quantity $|e\cos(\omega)|$, where $e$ is the orbital eccentricity and $\omega$ is the argument of periastron, see e.g., Charbonneau et al. (2005). For each planet, we take the average of our measured central phases, weighted by the inverse of their variance, deriving secondary eclipse central phases of $0.50053 \pm 0.00075$ (XO-

4b), 0.49983±0.00049 (HAT-P-8b) and 0.49897±0.00044 (HAT-P-6b).

The secondary eclipse central phases for XO-4b, HAT-P-6b and HAT-P-8b are close to the value of about 0.5, which is consistent with circular orbits, as expected from planets with orbital periods of a few days. The central phase of the secondary eclipse is not expected to be exactly 0.5, even for perfectly circular orbits, since it is affected by the light-travel time effect. The delays due to light-travel time are $55.09 \pm 0.30$, $44.81 \pm 0.50$ and $52.25 \pm 0.87$ seconds for XO-4b, HAT-P-8b and HAT-P-6b, respectively, assuming zero eccentricity. In addition, the day side of the planet is not expected to have uniform brightness, with the hottest spot located on the trailing side of the planet with respect to the planetary motion (e.g., Charbonneau et al. 2005; Cooper & Showman 2005; Knutson et al. 2007; Agol et al. 2010). This causes an additional apparent delay in the secondary eclipse. Scaling the results on HD 189733b by Agol et al. (2010), we estimate the expected delay due to a trailing planet hot spot to be about 40 s, 43 s and 49 s for XO-4b, HAT-P-8b and HAT-P-6b, respectively. These two combined effects suggest expected central eclipse phases of 0.50027 (XO-4b), 0.50033 (HAT-P-8b) and 0.50030 (HAT-P-6b). The fact that the average observed central phase for XO-4b is so close to the expected value is likely a coincidence, but suggests that the planet's orbit has extremely low eccentricity, if any at all. Another factor that can lead to secondary eclipses occurring earlier or later than expected are orbital perturbations by additional bodies in the system. In this case, different eclipses might be offset by varying amounts.

The secondary eclipse central phases of the three planets agree within $1.2\sigma$ (the two XO-4b eclipses), $0.3\sigma$ (HAT-P-6b) and $0.7\sigma$ (HAT-P-8b). Using the average of the times of secondary eclipse central phases, weighted by the inverse of the central phase variances, we can apply the equations on orbital eccentricity in Charbonneau et al. (2005). We find that, for HAT-P-8b, $|e\cos(\omega)| < 0.003$, while HAT-P-6b and XO-4b have $|e\cos(\omega)| < 0.004$, all three to the $3\sigma$ level. The limit for HAT-P-6b is in agreement with a measurement by Noyes et al. (2008) who found that the HAT-P-6b orbit is near circular, with $e = 0.046 \pm 0.031$. Preliminary results from more recent data, which is still under analysis, support this conclusion and indicate lack of significant orbital eccentricity (A. Howard, private communication).

HAT-P-6b has an inclined retrograde orbit, with a sky projected angle between the stellar and orbital axes, $\lambda = 166° \pm 10°$ (Hébrard et al. 2011). It is possible that planets on retrograde orbits may circularize somewhat faster than planets on prograde orbits (B. Jackson, private communication; Barker & Ogilvie 2009). Therefore, we do not expect that the retrograde orbit will have an effect on the eclipse central phase. The driving effects behind orbit circularization are the lags of the tidal bulges raised on both the star and the planet near periastron. The lag of the tide raised on the star by the planet can vary, depending on the speed and direction of the stellar rotation, relative to the planet, which will have an effect on the time required to circularize the planetary orbit.

## 5. CONCLUSION

Using *Spitzer/IRAC*, we measure eclipse depths at 3.6 and $4.5\,\mu$m for three planets, XO-4b, HAT-P-6b and HAT-P-8b. Our results are qualitatively the same regardless of the precise nature of the decorrelation function used for correcting the intra-pixel sensitivity variation in these Spitzer data. We find that XO-4b has an inverted atmosphere, HAT-P-6b's atmosphere is either weakly inverted or not inverted at all, while HAT-P-8b has no temperature inversion. Our measurements are not inconsistent with the studies of Knutson et al. (2010) and Cowan & Agol (2011) on the relationship between stellar chromospheric activity and temperature inversions, and planetary effective temperature and heat redistribution, respectively. We have improved the available ephemerides for our targets and we have put upper limits on their $e\cos(\omega)$ terms suggesting highly circularized orbits.

This work is based on observations made with the Spitzer Space Telescope, which is operated by the Jet Propulsion Laboratory, California Institute of Technology under a contract with NASA. Support for this work was provided by NASA through an award issued by JPL/Caltech. We thank the anonymous referee for a careful review of this paper.

TABLE 1
Adopted Stellar and Planetary Parameters

|  | XO-4b | HAT-P-8b[e] | HAT-P-6b[f] |
|---|---|---|---|
| $M_\star$ ($M_\odot$) | $1.32 \pm 0.02$[b] | $1.28 \pm 0.04$ | $1.29 \pm 0.06$ |
| $R_\star$ ($R_\odot$) | $1.56 \pm 0.05$[b] | $1.58^{+0.08}_{-0.06}$ | $1.46 \pm 0.06$ |
| $K_s$ (mag)[a] | $9.406 \pm 0.023$ | $8.953 \pm 0.013$ | $9.313 \pm 0.019$ |
| $T_{eff}$ (K) | $6400 \pm 70$[b] | 6130 | 6410 |
| $b_{impact}$ | $0.16 \pm 0.08$[c] | $0.32^{+0.09}_{-0.19}$ | $0.602 \pm 0.030$ |
| $M_p$ ($M_J$) | $1.72 \pm 0.20$[b] | $1.52^{+0.18}_{-0.16}$ | $1.057 \pm 0.119$ |
| $R_p$ ($R_J$) | $1.34 \pm 0.05$[b] | $1.50^{+0.08}_{-0.06}$ | $1.330 \pm 0.061$ |
| P (days)[d] | $4.1250823 \pm 0.0000039$ | $3.0763402 \pm 0.0000015$ | $3.8530030 \pm 0.0000014$ |
| $a_p$ (AU) | $0.0552 \pm 0.0003$[c] | $0.0449 \pm 0.0005$ | $0.05235 \pm 0.00087$ |

[a] Two Micron All Sky Survey (2MASS) $K_s$ magnitude of the star.
[b] McCullough et al. (2008).
[c] Narita et al. (2010).
[d] The orbital periods are taken from our updated ephemerides shown in Table 3.
[e] Values from Latham et al. (2009), except for the effective temperature, $T_{eff}$ (Knutson et al. 2010), and the period, P.
[f] Values from Noyes et al. (2008), except for the effective temperature, $T_{eff}$ (Knutson et al. 2010), and the period, P.

TABLE 2
Observation Details

|  | XO-4b | | HAT-P-8b | | HAT-P-6b | |
|---|---|---|---|---|---|---|
|  | 3.6 $\mu$m | 4.5 $\mu$m | 3.6 $\mu$m | 4.5 $\mu$m | 3.6 $\mu$m | 4.5 $\mu$m |
| Observation start (UTC) | 15-Dec-09,7:26 | 7-Dec-09,1:28 | 14-Jan-10,16:18 | 11-Jan-10,14:35 | 11-Sep-10,23:21 | 19-Sep-10,16:14 |
| Observation end (UTC) | 15-Dec-09,15:18 | 7-Dec-09,9:21 | 14-Jan-10,24:00 | 11-Jan-10,22:17 | 12-Sep-10,7:02 | 19-Sep-10,23:55 |
| Phase span | 0.452 to 0.532 | 0.453 to 0.533 | 0.437 to 0.542 | 0.439 to 0.543 | 0.453 to 0.537 | 0.453 to 0.536 |
| Image count | 14,080 | 14,080 | 13,760 | 13,760 | 13,760 | 13,760 |

TABLE 3
Ephemerides

| N | Observation Date | Primary Transit (BJD$_{TDB}$) | Uncertainty | Notes |
|---|---|---|---|---|
| **XO-4b** | | | | |
| 0 | 2008-01-20 | 2454485.93297 | 0.00040 | McCullough et al. (2008) |
| 63 | 2008-10-06 | 2454745.81494 | 0.00105 | Narita et al. (2010), egress only |
| 64 | 2008-10-10 | 2454749.93969 | 0.00114 | Narita et al. (2010), ingress only |
| 79 | 2008-12-11 | 2454811.81238 | 0.00069 | UDEM, $I_c$ band |
| 86 | 2009-01-09 | 2454840.68979 | 0.00082 | UDEM, $I_c$ band |
| 87 | 2009-01-13 | 2454844.81636 | 0.00063 | UDEM, $I_c$ band |
| 144 | 2009-09-05 | 2455079.94672 | 0.00110 | Narita et al. (2010), ingress only |
| 182 | 2010-02-09 | 2455236.69753 | 0.00065 | Narita et al. (2010) |

These eight transits yield:

P = 4.1250823 ± 0.0000039 days

$T_0$ = 2454485.93307 ± 0.00036 in BJD$_{TDB}$

| **HAT-P-8b** | | | | |
|---|---|---|---|---|
| 0 | 2007-12-03 | 2454437.67657 | 0.00034 | Latham et al. (2009) |
| 316 | 2010-08-01 | 2455409.80055 | 0.00063 | UDEM, $I_c$ band |
| 325 | 2010-08-28 | 2455437.48698 | 0.00039 | TRESCA database, R band[a] |

These three transits yield:

P = 3.0763402 ± 0.0000015 days

$T_0$ = 2454437.67657 ± 0.00034 in BJD$_{TDB}$

| **HAT-P-6b** | | | | |
|---|---|---|---|---|
| 0 | 2006-10-27 | 2454035.67648 | 0.00028 | Noyes et al. (2008); Szabo et al. (2010) |
| 81 | 2007-09-04 | 2454347.76839 | 0.00042 | Noyes et al. (2008); Szabo et al. (2010) |
| 172 | 2008-08-19 | 2454698.3916 | 0.0011 | Szabo et al. (2010) |
| 183 | 2008-10-01 | 2454740.77668 | 0.00063 | UDEM, $I_c$ band |
| 292 | 2009-11-25 | 2455160.75292 | 0.00034 | KPNO 2.1 m, J band |
| 362 | 2010-08-21 | 2455430.4657 | 0.0013 | TRESCA database, R band[b] |

These six transits yield:

P = 3.8530030 ± 0.0000014 days

$T_0$ = 2454035.67617 ± 0.00025 in BJD$_{TDB}$

[a] Calar Alto Observatory, Spain, 1.2m telescope, observer: Felix Hormuth, published in the TRESCA database, http://var2.astro.cz/EN/tresca/
[b] Montcabrer MPC213, Cabrilis, Spain, 12″ telescope, observer: Ramon Naves, published in the TRESCA database, http://var2.astro.cz/EN/tresca/

TABLE 4
Secondary Eclipse Results

| | Eclipse Depth (%) | Eclipse Central Phase | $\mathrm{BJD_{TDB}}^a - 2\,400\,000$ | O − C[b] (min) |
|---|---|---|---|---|
| XO-4b, 3.6 $\mu$m | $0.056^{+0.012}_{-0.006}$ | $0.50197^{+0.00164}_{-0.00138}$ | $55\,181.01756^{+0.00673}_{-0.00565}$ | $10.10^{+9.74}_{-8.20}$ |
| XO-4b, 4.5 $\mu$m | $0.135^{+0.010}_{-0.007}$ | $0.50005^{+0.00087}_{-0.00083}$ | $55\,172.75948^{+0.00186}_{-0.00144}$ | $-1.31^{+5.17}_{-4.93}$ |
| HAT-P-8b, 3.6 $\mu$m | $0.131^{+0.007}_{-0.010}$ | $0.49967^{+0.00033}_{-0.00077}$ | $55\,211.37512^{+0.00089}_{-0.00231}$ | $-2.92^{+1.46}_{-3.41}$ |
| HAT-P-8b, 4.5 $\mu$m | $0.111^{+0.008}_{-0.007}$ | $0.50042^{+0.00101}_{-0.00107}$ | $55\,208.30108^{+0.00234}_{-0.00258}$ | $0.40^{+4.47}_{-4.74}$ |
| HAT-P-6b, 3.6 $\mu$m | $0.117 \pm 0.008$ | $0.49931^{+0.00089}_{-0.00123}$ | $55\,451.65211^{+0.00339}_{-0.00470}$ | $-5.49^{+4.94}_{-6.82}$ |
| HAT-P-6b, 4.5 $\mu$m | $0.106 \pm 0.006$ | $0.49890^{+0.00029}_{-0.00067}$ | $55\,459.35654^{+0.00096}_{-0.00250}$ | $-7.77^{+1.61}_{-3.72}$ |

[a] Time of secondary eclipse central phase, in Barycentric Julian Date (BJD) based on barycentric dynamic time (TDB).
[b] The measured offset from the expected central phase as described in Section 4.2, in minutes.

TABLE 5
Planetary Temperatures

| Planet | Wavelength | $T_b(\lambda)$[a] in K | $T_d$[b] in K | $T_{\varepsilon=0}$[c] in K | $T_d/T_0$[d] |
|---|---|---|---|---|---|
| XO-4b | 3.6 $\mu$m | $1522 \pm 92$ | $1577 \pm 106$ | $2084 \pm 27$ | $0.68 \pm 0.05$ |
| | 4.5 $\mu$m | $1961 \pm 65$ | | | |
| HAT-P-8b | 3.6 $\mu$m | $1923 \pm 56$ | $1900 \pm 108$ | $2243 \pm 52$ | $0.77 \pm 0.05$ |
| | 4.5 $\mu$m | $1588 \pm 48$ | | | |
| HAT-P-6b | 3.6 $\mu$m | $1934 \pm 91$ | $1913 \pm 128$ | $2087 \pm 42$ | $0.83 \pm 0.06$ |
| | 4.5 $\mu$m | $1657 \pm 41$ | | | |

[a] Brightness temperature.
[b] Effective dayside temperature.
[c] Effective temperature of the planet's day side assuming no heat redistribution to the night side.
[d] Ratio of the effective dayside temperature ($T_d$) and the equilibrium temperature of the planet at the substellar point ($T_0$).

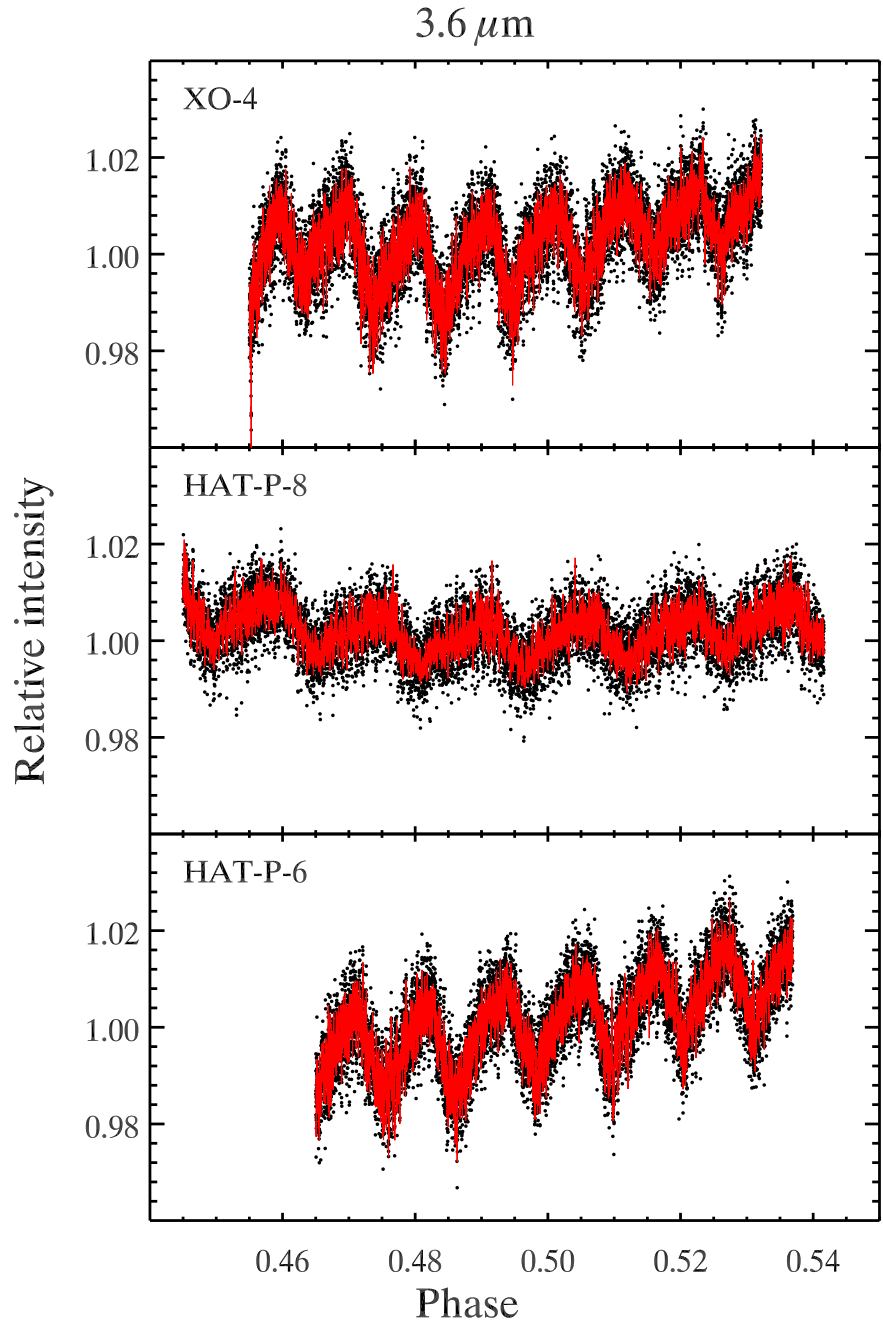

FIG. 1.— In this figure we present the uncorrected time series photometry for XO-4, HAT-P-6 and HAT-P-8 at 3.6 μm during secondary eclipse (dots). The imposed red lines represent the best fit eclipse models we obtain, which include the eclipse itself, a linear ramp and the dependence of the measured intensity on the x and y-position of the stellar PSF centroid on the detector array discussed in Section 3.2. All photometric points used in the analysis are shown here, but not any rejected data.

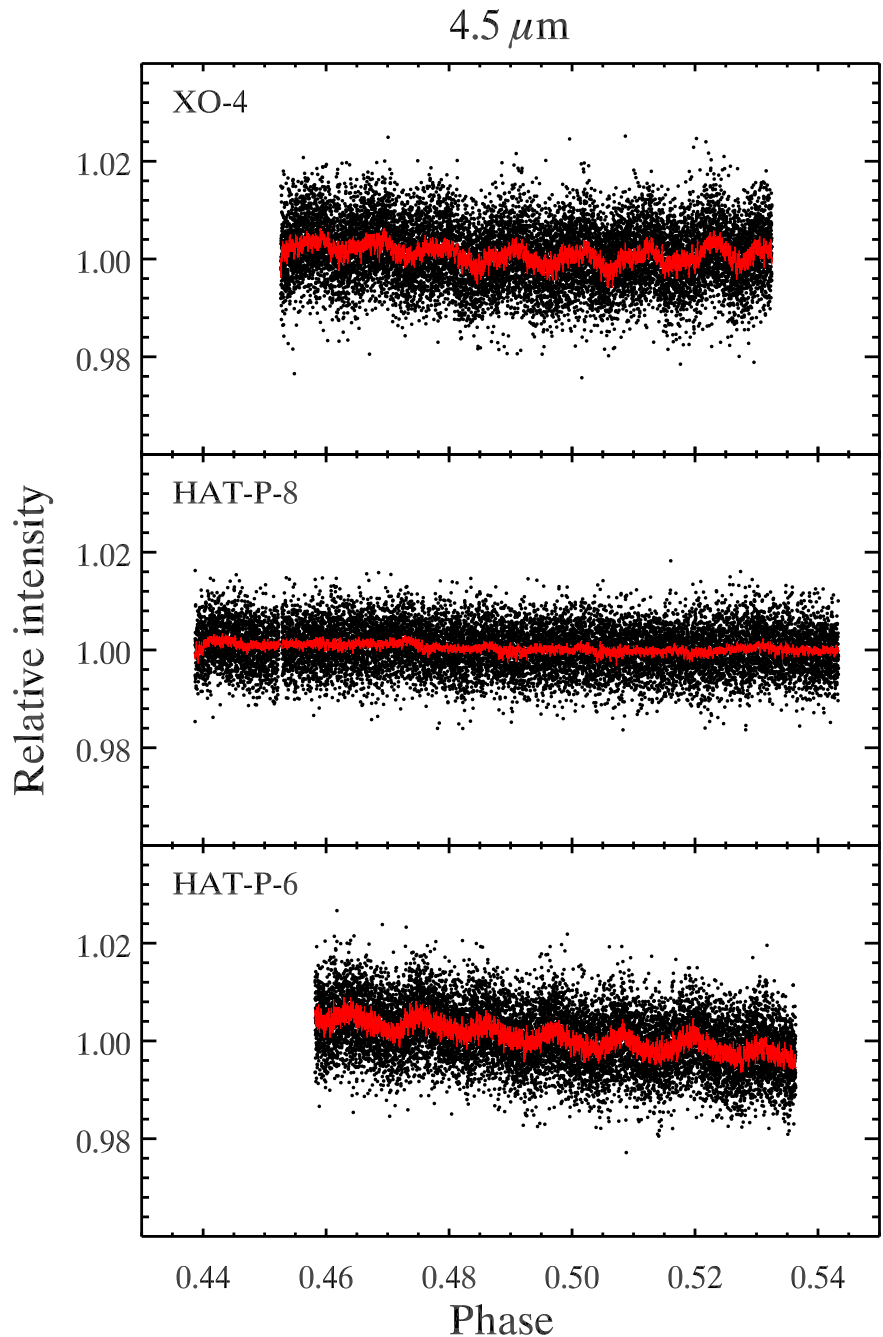

FIG. 2.— Similar to Figure 1, but in this plot we show the 4.5 $\mu$m data. The HAT-P-8 raw light curve in 4.5 $\mu$m shows very little variation due to the unusually small drift of the telescope pointing – by a factor of about 1.5 less than the HAT-P-6 data, and about 2 less than the XO-4 data in this wavelength. Throughout the whole duration of the observation, the pointing varied by about 0.15 pixels in X and about 0.10 pixels in Y. For comparison, during the XO-4 observation at 4.5 $\mu$m the telescope pointing drifted within a range of $\sim$ 0.3 pixels in X and $\sim$ 0.3 pixels in Y, including several changes in the pixel on which the star was centered. In addition, in the HAT-P-8 and HAT-P-6 observations, when the telescope pointing moved away from the center of the pixel along one axis, it moved closer to the center along the other axis, in effect canceling the decrease of sensitivity effect, as noted previously by Anderson et al. (2011) in their Spitzer secondary eclipse data on WASP-17b. In the XO-4 data, the telescope pointing drifted toward or away from the center of the pixel simultaneously along both axes, enhancing the apparent sensitivity variation.

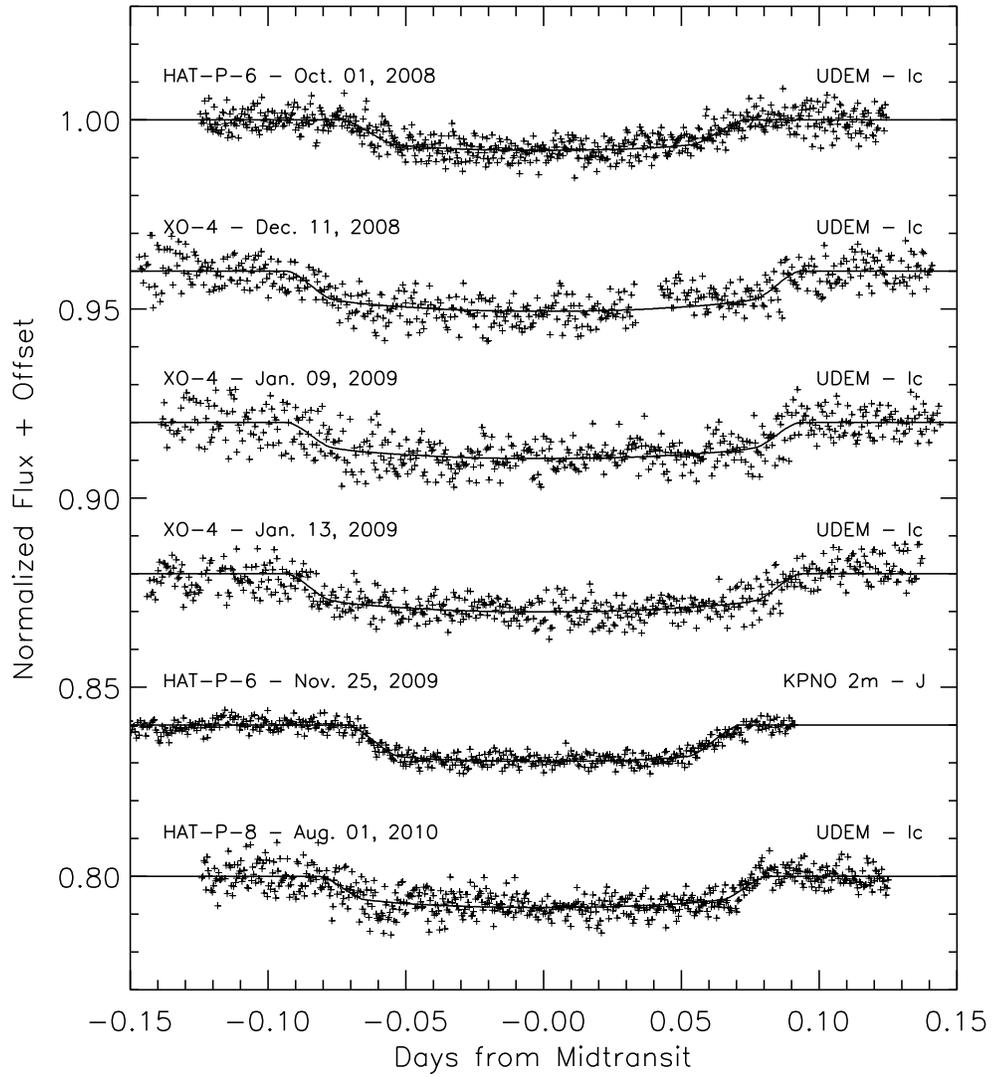

FIG. 3.— The transit curves of XO-4b, HAT-P-6b and HAT-P-8b were observed by the Universidad de Monterrey Observatory in the $I_c$ band and the KPNO 2.1 m telescope in the J band. We fit model transit curves (solid lines) to the data in order to estimate the central transit time, and improve the ephemeris for the three planets.

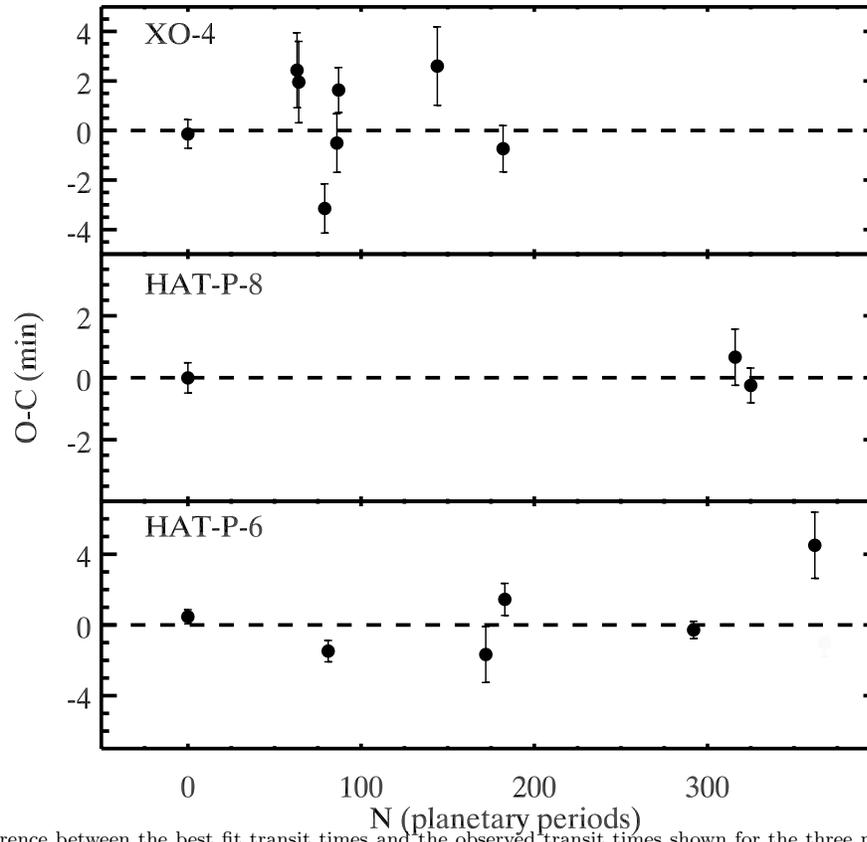

FIG. 4.— The difference between the best fit transit times and the observed transit times shown for the three planets. The horizontal axis represents the number of periods after the $T_0$ transit. We do not see any significant residuals in the timing data that could signify transit timing variations. The three observed HAT-P-8b transits have the smallest residuals, but this may be due to the small sample size.

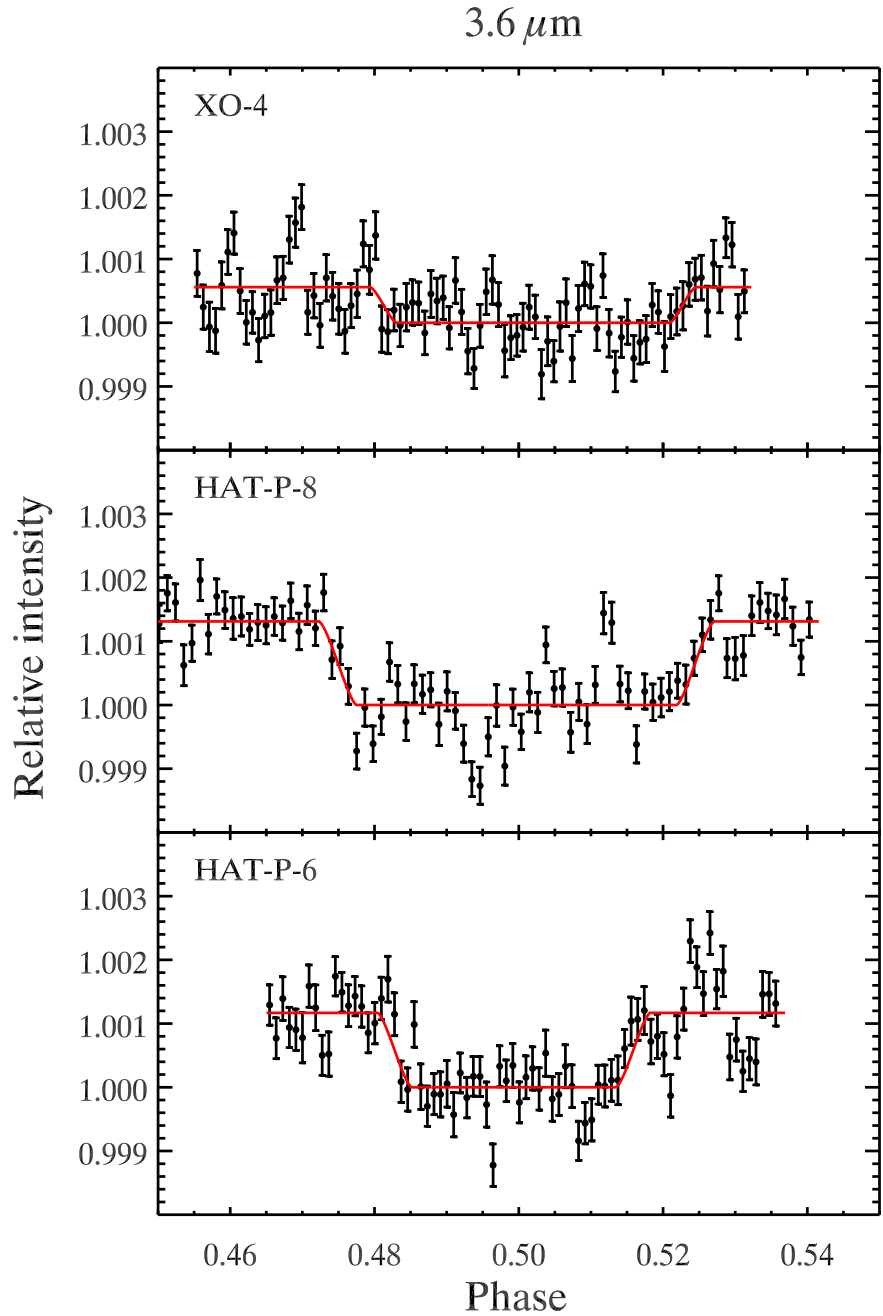

FIG. 5.— The secondary eclipse photometry at $3.6\,\mu$m shown after corrections for instrumental effects. The data points are binned, with bin size of 150 (about 5 min 3 sec), where the error bars represent the standard deviation of the points within the bin, divided by the square root of their number. The red lines represent the best fit secondary eclipse model. In order to focus better on the secondary eclipse, we have omitted data points with phase < 0.45 from this plot (HAT-P-8b only).

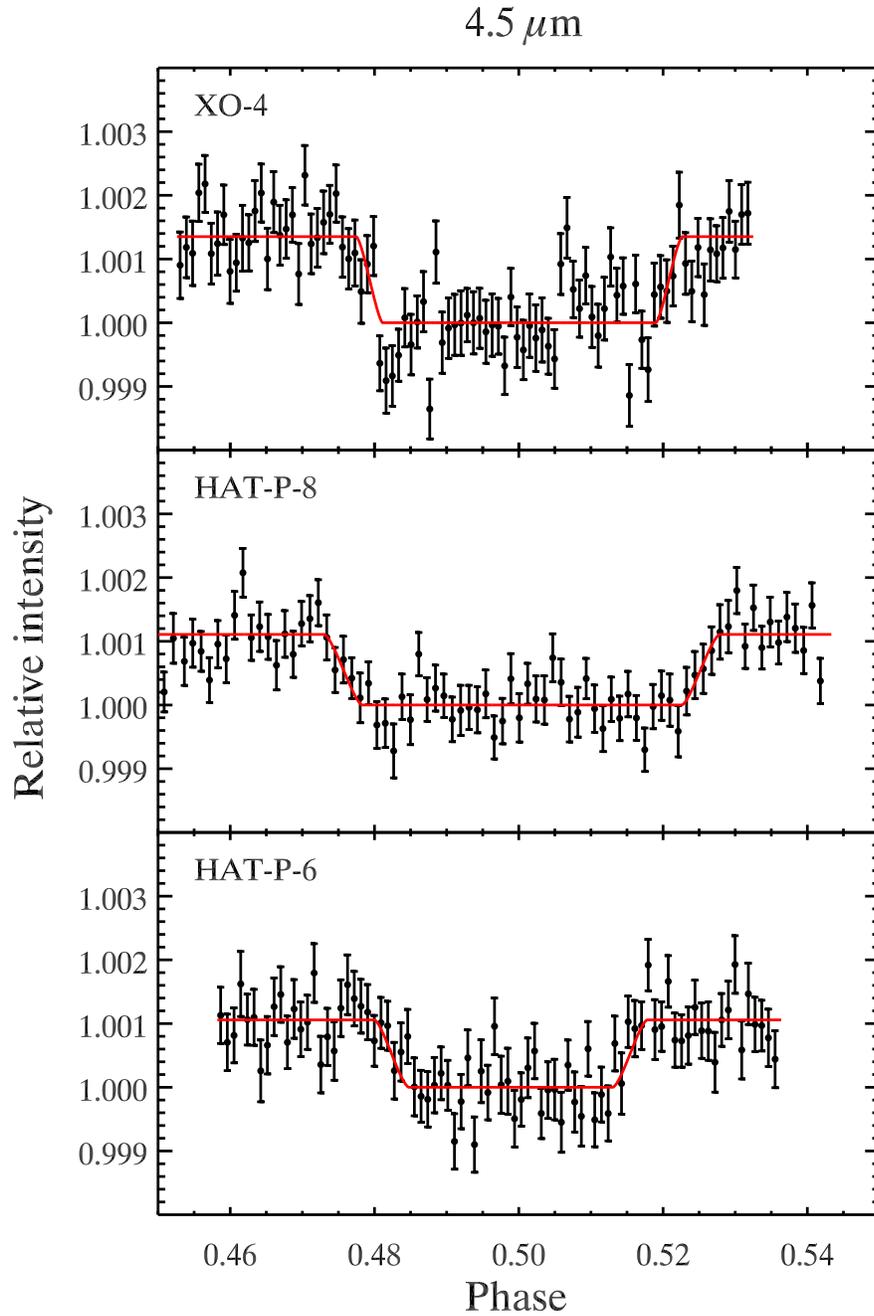

FIG. 6.— Similar to Figure 5, but for the 4.5 $\mu$m secondary transit light curves. Here, the bins also contain 150 points, but their coverage in time is about 5 min 10 sec.

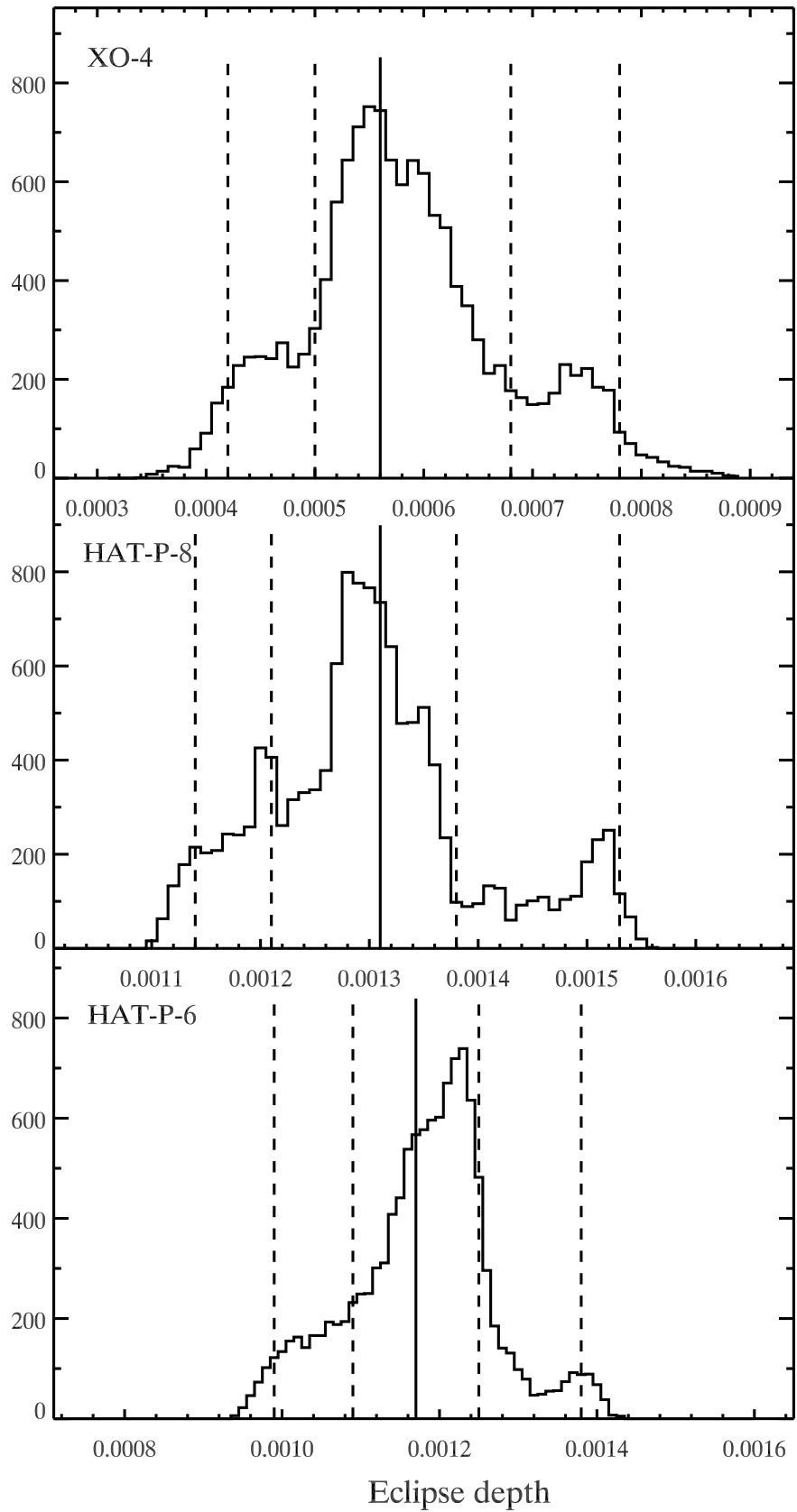

Fig. 7.— The histograms of the eclipse depths that result from the "prayer bead" simulated data fits of the three occultations at 3.6 $\mu$m (see Section 3.2.1) have non-Gaussian distributions. We calculate the regions centered on the best fit result (solid line) from the original data, that contain 68% (1$\sigma$) and 95% (2$\sigma$) of the eclipse depths derived from the simulated data sets (inner and outer dashed lines).

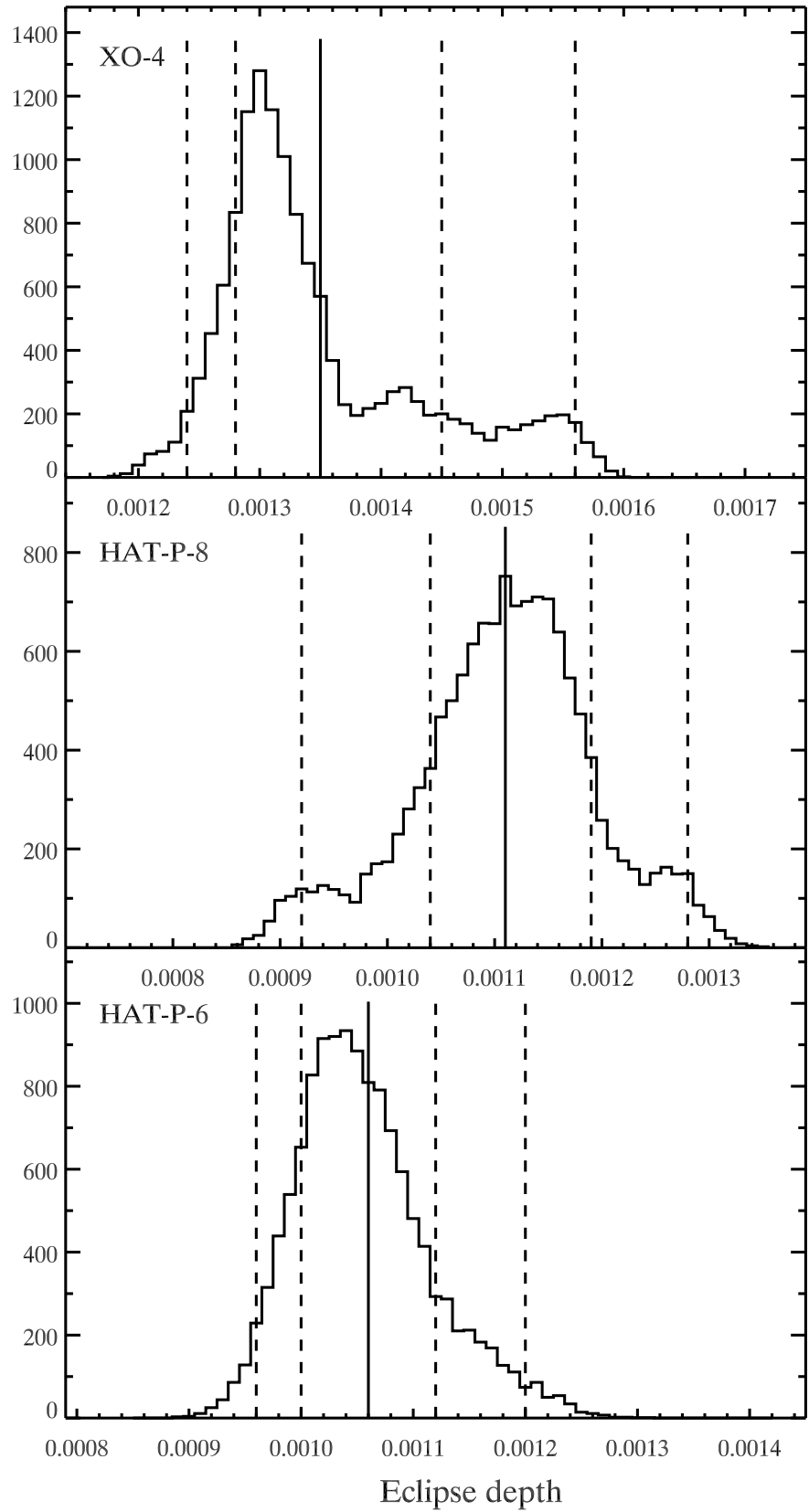

Fig. 8.— These histograms are similar to the ones in Figure 7, but show the eclipse depths at 4.5 μm simulated via the "prayer bead" method.

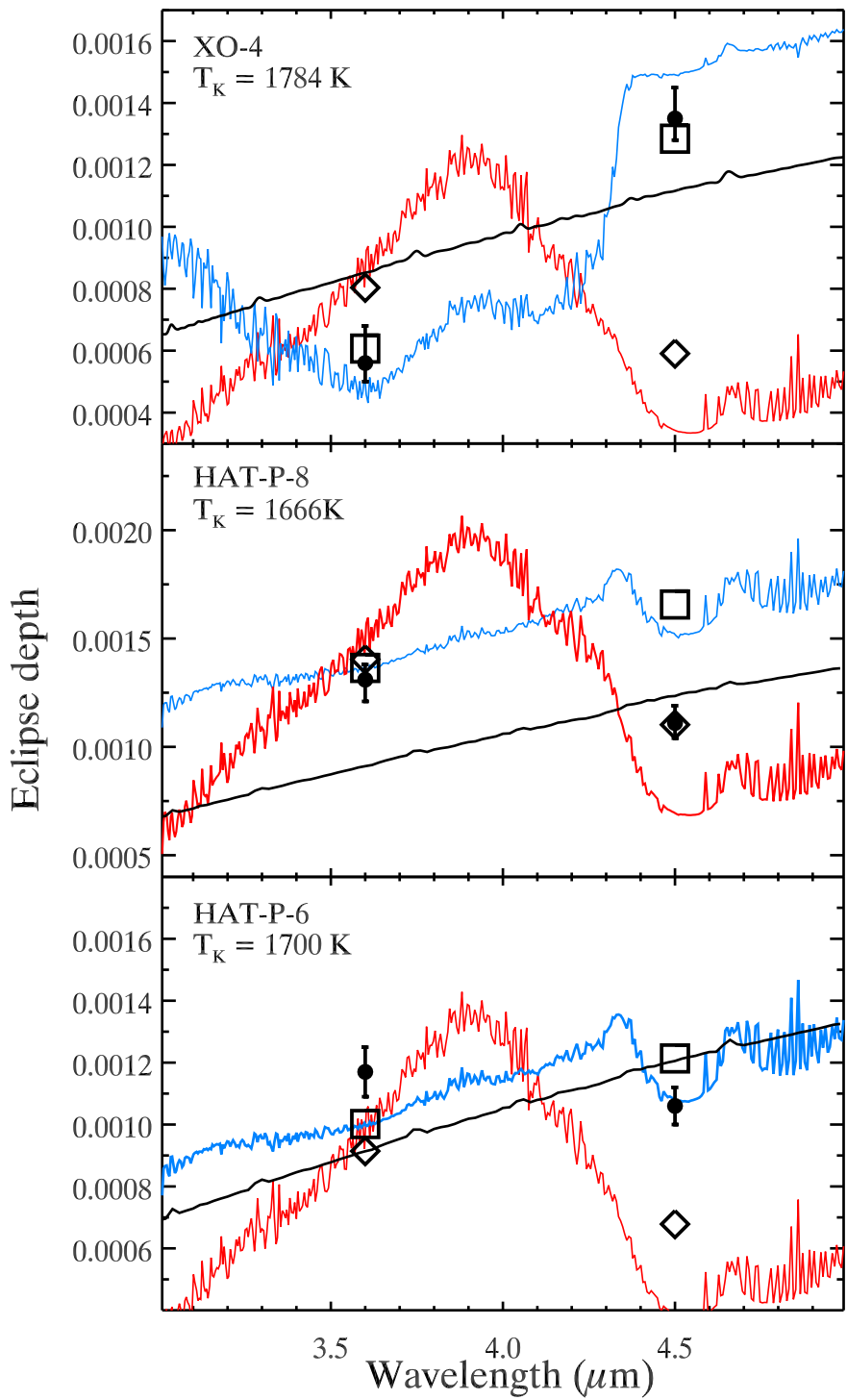

Fig. 9.— The measured secondary eclipse amplitudes (filled circles), compared with a best fit model with black body planet and a Kurucz model for the star (black line; Kurucz 1979). The corresponding black body temperatures, $T_K$, for the planets are labeled on the plots. We have assumed the stellar effective temperatures listed in Knutson et al. (2010). In blue, we show the inverted Burrows models, and the red lines represent solar composition non-inverted Burrows models (Burrows et al. 2007, 2008). We over-plot the expected contrast based on integrating the model stellar planetary fluxes over the IRAC pass-bands (diamonds for the non-inverted models and squares for the inverted models). Based on the Burrows models, the measured XO-4b eclipse depths can be explained by a strong temperature inversion in the upper layers of its atmosphere and moderately efficient redistribution of heat to its night side ($\kappa_{abs} = 0.4\,\mathrm{cm}^2\mathrm{g}^{-1}$, $P_n = 0.35$; Burrows et al. 2007, 2008). The HAT-P-6b transit amplitudes imply a very moderate temperature inversion ($\kappa_{abs} = 0.1\,\mathrm{cm}^2\mathrm{g}^{-1}$), and can be matched by a model with $P_n = 0.3$. The HAT-P-8b results are explained best by a model with no inversion at all and inefficient heat redistribution ($\kappa_{abs} = 0\,\mathrm{cm}^2\mathrm{g}^{-1}$, $P_n = 0.1$).

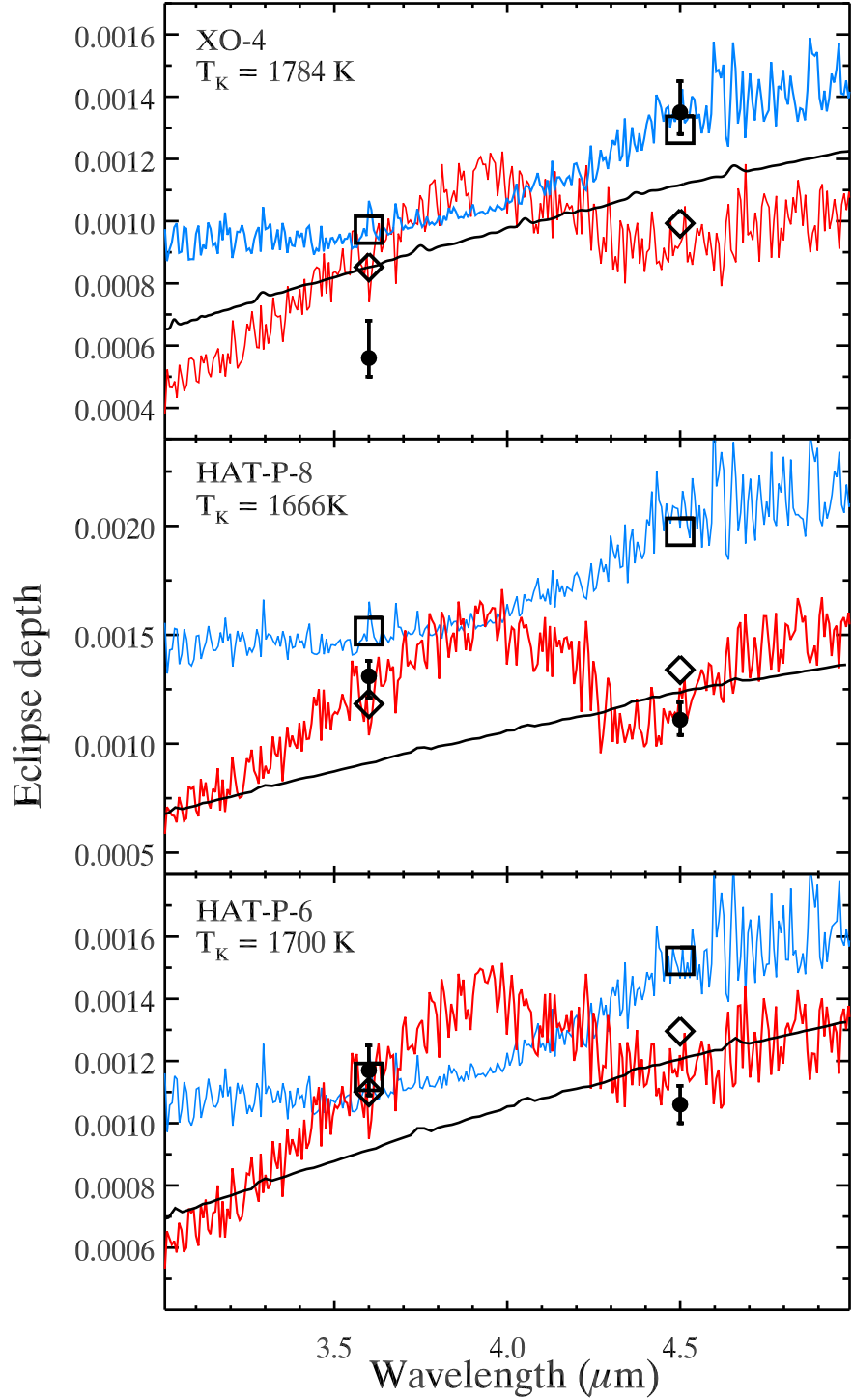

FIG. 10.— Similar to Figure 9, but here our measurements are compared to models by Fortney et al. (2005, 2006a, 2008). Again, the inverted models are represented with squares and blue lines, and the non-inverted models are shown with diamonds and red lines. The models displayed here have solar metallicities, except the non-inverted one for HAT-P-8b, which has metallicity that is 10× solar. For the inverted XO-4b model, best matching our eclipse depth measurements, the factor $f = 0.5$, which means that the incident flux is redistributed uniformly over the dayside, but none is redistributed to the night side. TiO and VO are present in the atmosphere, causing the temperature inversion. The non-inverted HAT-P-8b model, which best explains the measurements, also has $f = 0.5$, but the TiO and VO are removed from the atmosphere, so there is no temperature inversion. For HAT-P-6b, a non-inverted model with $f = 0.63$ corresponds to the measured eclipse depths best. In this parametrization, $f = 0.67$ is the maximum value, meaning that no incident flux is redistributed at all, not even within the dayside hemisphere. TiO and VO are not present in the atmosphere of HAT-P-6b in this model, suggesting an atmosphere with a non-inverted temperature profile.

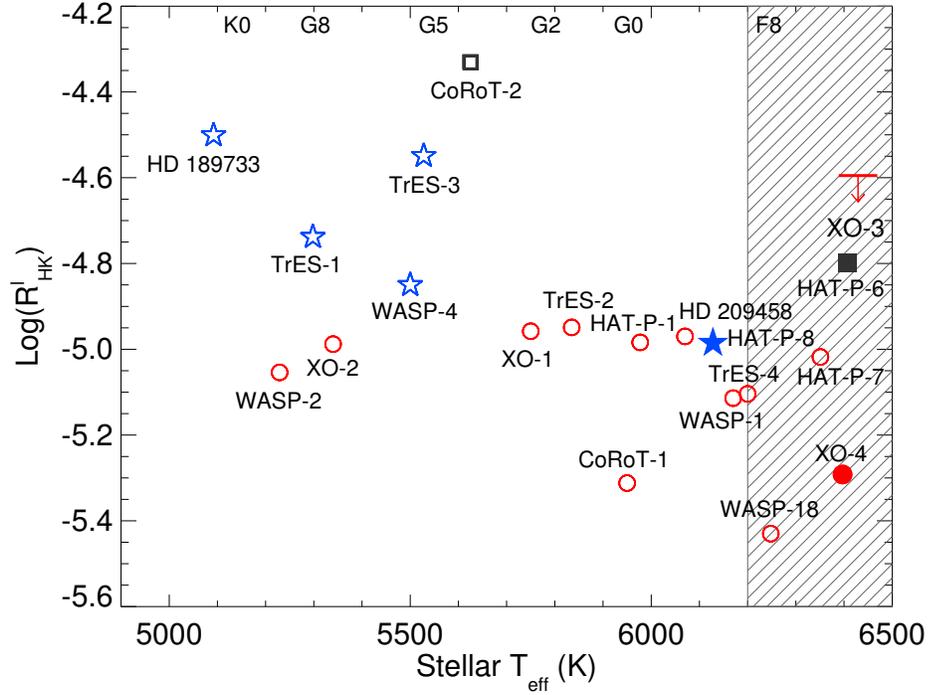

FIG. 11.— This plot of stellar Ca II H & K activity index versus effective temperature of the host star is taken from Knutson et al. (2010), but we have added the points from our study (filled symbols). The planets with temperature inversions in their upper atmospheric layers (red circles) seem to be grouped lower in this plot than the non-inverted planets (blue stars). The shaded area represents the temperature range where the chromospheric activity measurements using the Ca II H & K lines are not well calibrated (Noyes et al. 1984). The CoRoT-2 (empty square) planet has secondary eclipse depths that are poorly matched both by inverted and non-inverted models (Deming et al. 2011), while HAT-P-6b (filled square) may have either a weak temperature inversion or no inversion at all. XO-3 (red downward arrow) has a planet with an inverted atmosphere, but its host star activity index has only been assigned an upper limit by Knutson et al. (2010).